\documentclass[useAMS,usenatbib]{mn2e}
\usepackage{mn2e-breakabs}
\usepackage{graphicx}
\usepackage{times}
\def\la{\mathrel{\mathpalette\fun <}}

\def\fun#1#2{\lower3.6pt\vbox{\baselineskip0pt\lineskip.9pt
        \ialign{$\mathsurround=0pt#1\hfill##\hfil$\crcr#2\crcr\sim\crcr}}}

\def\bfk{\mbox{\bf k}}
\def\bfr{\mbox{\bf r}}

\def\rd{{\rm d}}

\newcommand{\be}{\begin{equation}}
\newcommand{\ee}{\end{equation}}
\newcommand{\ba}{\begin{eqnarray}}
\newcommand{\ea}{\end{eqnarray}}
\newcommand{\simgt}{\,\hbox{\lower0.6ex\hbox{$\sim$}\llap{\raise0.6ex\hbox{$>$}}}\,}
\newcommand{\simlt}{\,\hbox{\lower0.6ex\hbox{$\sim$}\llap{\raise0.6ex\hbox{$<$}}}\,}

\begin{document}

\title{Toward More Realistic Forecasting of Dark Energy Constraints from Galaxy Redshift Surveys}

\author[Wang, Chuang, \& Hirata]{
  \parbox{\textwidth}{
    Yun Wang\thanks{E-mail: wang@nhn.ou.edu}$^1$, Chia-Hsun Chuang\thanks{MultiDark Fellow}$^2$, \& Christopher M. Hirata$^3$}
  \vspace*{4pt} \\
  $^1$ Homer L. Dodge Department of Physics \& Astronomy, Univ. of Oklahoma,
                 440 W Brooks St., Norman, OK 73019, U.S.A.\\
  $^2$ Instituto de F\'{\i}sica Te\'orica, (UAM/CSIC), Universidad Aut\'onoma de Madrid,  Cantoblanco, E-28049 Madrid, Spain \\
  $^3$ Caltech M/C 350-17, Pasadena, CA 91125, U.S.A.
                 }

\date{\today}

\maketitle

\begin{abstract}

Galaxy redshift surveys are becoming increasingly important as a dark energy probe.
We improve the forecasting of dark energy constraints from galaxy redshift surveys
by using the ``dewiggled'' galaxy power spectrum, $P_{dw}(\bfk)$, in the Fisher
matrix calculations.
Since $P_{dw}(\bfk)$ is a good fit to real galaxy clustering data over most
of the scale range of interest, our approach is more realistic compared to
previous work in forecasting dark energy constraints from galaxy redshift surveys.
We find that our new approach gives results in excellent agreement when compared
to the results from the actual data analysis of the clustering of the 
Sloan Digital Sky Survey DR7 luminous red galaxies.
We provide forecasts of the dark energy constraints from a plausible
Stage IV galaxy redshift survey.


\end{abstract}


\section{Introduction}  \label{sec:intro}

One of the most important discoveries in modern cosmology is the accelerated expansion of the Universe \citep{Riess98,Perl99}.
The power spectrum or 2-point correlation function measured from galaxy redshift surveys has provided one of the primary probes of
cosmic acceleration, both through the broadband measurement of the shape imprinted by matter-radiation equality \citep[e.g.][]{Percival01, Tegmark04} and
through the baryon-acoustic oscillation (BAO) feature imprinted at recombination \citep[e.g.][]{Eisenstein05}.
Galaxy clustering also allows us to differentiate smooth
dark energy and modified gravity as the cause for cosmic acceleration
through the simultaneous measurements of the cosmic expansion history
$H(z)$, and the growth rate of cosmic large scale structure, $f_g(z)$ \citep{Guzzo08,Wang08a,Blake12}.

The Fisher matrix approach has generally been used in the forecasts
of future galaxy redshift surveys. In this paper, we improve the Fisher matrix approach 
by making it more realistic. This enables its use in cross-checking dark energy
and gravity constraints from current galaxy clustering data, as well as in
making the forecasts for future galaxy redshift surveys more robust and
reliable.

We present our method in Section~\ref{sec:method}, our results in Section~\ref{sec:results}, and
summarize and conclude in Section~\ref{sec:conclusion}.

\section{Method}
\label{sec:method}

The redshift-space galaxy power spectrum $P(k,\mu)$ is a rich source of cosmological information. It includes the BAO feature \citep{BG03,SE03}, which has received a great deal of attention as a standard ruler that can be used in both the transverse direction (to measure distances) and the radial direction (to measure the Hubble rate). However, the full galaxy power spectrum at large scales is also sensitive to the underlying matter power spectrum, to the growth of structure via redshift-space distortions \citep{Kaiser87}, and to standard ruler effects. This additional information requires some work to extract, since one must simultaneously measure the cosmology and the galaxy biasing parameters. Nevertheless, the galaxy power spectrum provides the most powerful constraints on dark energy and gravity. In this paper, we focus on the analysis of 
the full set of 2-point galaxy statistics, and do {\em not} limit ourselves to {\em only} the BAO information.

\subsection{Formalism}

Our Fisher matrix approach is derived from that of \cite{SE03},
and based on \cite{Wang06,Wang08a,Wang10} and \cite{Wang_etal10}.
In the limit where the length scale corresponding to the survey volume is 
much larger than the scale of any features in the observed galaxy power spectrum
$P_g(\bfk)$, we can assume that the likelihood function for the band powers of a galaxy 
redshift survey is Gaussian \citep{FKP}, with a measurement error
in $\ln P(\bfk)$ that is proportional to $[V_{\rm eff}(\bfk)]^{-1/2}$, 
with the effective volume of the survey defined as
\ba
V_{\rm eff}(k,\mu)&\equiv& \int d\bfr^3 \left[ \frac{n(\bfr) P_g(k,\mu)}
{ n(\bfr) P_g(k,\mu)+1} \right]^2\nonumber\\
&=&\left[ \frac{ n P_g(k,\mu)}{n P_g(k,\mu)+1} \right]^2 V_{\rm survey},
\ea
where the comoving number density $n$ is assumed to only depend on
the redshift (and constant in each redshift slice) for simplicity
in the last part of the equation.

In order to propagate the measurement error in $\ln P_g(\bfk)$ 
into measurement errors for the parameters $p_i$, we use
the Fisher matrix \citep{Tegmark97}
\be
F_{ij}= \int_{k_{\rm min}}^{k_{\rm max}}
\frac{\partial\ln P_g(\bfk)}{\partial p_i}
\frac{\partial\ln P_g(\bfk)}{\partial p_j}\,
V_{\rm eff}(\bfk)\, \frac{\rd \bfk^3}{2\, (2\pi)^3},
\label{eq:Fisher_full}
\ee
where $p_i$ are the parameters to be estimated from data, and 
the derivatives are evaluated at parameter values of the
fiducial model. Note that the Fisher matrix $F_{ij}$ is the 
inverse of the covariance matrix of the parameters $p_i$ if 
the $p_i$ are Gaussian distributed.

We adopt the standard notation that $\bfk$ can be decomposed into a line-of-sight component $k_\parallel$ and the transverse or in-the-plane-of-the-sky component $k_\perp$. The cosine of the angle between $\bfk$ and the line of sight vector is denoted by $\mu = k_\parallel/|\bfk|$.

\subsection{The model for the galaxy power spectrum}

At cosmological distances, the ``true'' galaxy power spectrum is {\em not} a direct observable, since one can measure a galaxy's position only in angular and redshift coordinates and not in its true 3D comoving coordinates. This is of course the basis for extraction of the ``standard ruler'' information, including the \citet{AP79} effect. Therefore standard practice is to project the galaxies to their comoving positions assuming some reference cosmology (or fiducial cosmology), and then a power spectrum or correlation function estimator is applied. The observed galaxy power spectrum is then related to the true galaxy power spectrum via a coordinate transformation: the wavenumber $\bfk^{\rm ref}$ in the reference cosmology is related to the wavenumber in the true cosmology via
\begin{equation}
k^{\rm ref}_\perp = \frac{D_A(z)}{D_A^{\rm ref}(z)}k_\perp ~~{\rm and}~~
k^{\rm ref}_\parallel = \frac{H^{\rm ref}(z)}{H(z)} k_\parallel.
\end{equation}

Based on \cite{SE03} and \cite{CW11}, our model for $P_{g}(\bfk)$ can then be written as
\ba
P_{g}(k^{\rm ref}_{\perp},k^{\rm ref}_{\parallel}) &=&
\frac{\left[D_A^{\rm ref}(z)\right]^2  H(z)}{\left[D_A(z)\right]^2 H^{\rm ref}(z)}
\, b^2 \frac{\left( 1+\beta\, \mu^2 \right)^2}
{1+k^2 \mu^2 \sigma^2_{r,p}/2}
\nonumber\\
&& \times P_{\rm dw}(\bfk)_{z}\, e^{-k^2\mu^2 \sigma^2_{r,z}} + P_{\rm shot},
\label{eq:P(k)b}
\ea
where $H(z)=\dot{a}/{a}$ (with $a$ denoting the cosmic scale factor) is
the Hubble parameter, and $D_A(z)=r(z)/(1+z)$ is the angular diameter distance at $z$,
with the comoving distance $r(z)$ given by
\be
\label{eq:r(z)}
 r(z)=c\, |\Omega_k|^{-1/2} {\rm sinn}\left[|\Omega_k|^{1/2}\, 
\int_0^z\frac{dz'}{H(z')}\right],
\ee
where ${\rm sinn}(x)=\sin(x)$, $x$, $\sinh(x)$ for 
$\Omega_k<0$, $\Omega_k=0$, and $\Omega_k>0$ respectively.
In addition to the geometrical distortion, this model includes the linear galaxy bias and redshift-space distortion (RSD), nonlinear smearing of the BAO feature, halo shot noise, small-scale peculiar velocities, and redshift errors.

The bias between galaxy and matter distributions is denoted by
$b(z)$. The linear RSD parameter
$\beta(z)=f_g(z)/b(z)$ \citep{Kaiser87}, where $f_g(z)$ is the linear growth rate;
it is related to the linear growth factor $G(z)$ 
(normalized such that $G(0)=1$) as follows 
\be
f_g(z)=\frac{\mbox{d}\ln G(z)}{\mbox{d}\ln a}.
\ee
We have assume that the peculiar velocities of galaxies can be
modeled with a probability distribution
\be
f(v)=\frac{1}{\sigma_p \sqrt{2}} \, e^{-\sqrt{2}|v|/\sigma_p},
\label{eq:f(v)}
\ee
where $\sigma_p$ is the pairwise peculiar velocity dispersion.
The Fourier transform of $f(v)$ is $1/[1+k^2 \mu^2 \sigma^2_{r,p}/2]$, 
the small scale RSD factor included in Eq.~(\ref{eq:P(k)b}) \citep{Hamilton98}.
Note that $\sigma_{r,p}$ is the distance dispersion corresponding
to the physical velocity dispersion $\sigma_p$, thus
$\sigma_p=H(z) [a(z)\sigma_{r,p}]$, and
\be
\label{eq:sigma_{r,p}}
 \sigma_{r,p}=\frac{\sigma_p}{H(z) a(z)}.
\ee
Note that we have adopted minimal small scale RSD modeling in this work
(see Eq.[\ref{eq:f(v)}]), since we only consider quasilinear scales for a conservative approach.
The limitations of Eq.(\ref{eq:f(v)}) have been discussed in detail
by \cite{Scoccimarro04}. When smaller scales ($\la 20\,h^{-1}$Mpc) are included
in the analysis, it will be critical to use an improved RSD model,
see, e.g., \cite{CW12b}.

An additional damping factor, $e^{-k^2\mu^2\sigma_{r,z}^2}$, is 
inserted to account for redshift uncertainties, with $\sigma_{r,z} = (\partial r/\partial z) \sigma_z$. This is intended to incorporate the true redshift uncertainty resulting from fitting the centroid of an emission line (in an emission line survey), but this factor could also absorb other small errors in the redshift (e.g. due to the emission line velocity not being exactly equal to zero in the rest frame of the galaxy's host halo).

Nonlinear smearing of the BAO feature occurs due to the small-scale (i.e. $\ll s_{\rm BAO}\sim 150$ Mpc) displacements during structure formation. These displacements take sharp, coherent features in the correlation function at large scales (e.g. the BAO) and smear them out; in Fourier space, this corresponds to a damping of the oscillatory part of $P(\bfk)$.
This effect is modeled by using the dewiggled matter power spectrum at redshift $z$, given by
\be
P_{\rm dw}(\bfk,z)= G^2(z)P_0 k^{n_S} T^2_{\rm dw}(\bfk,z).
\label{eq:P(k|z)}
\ee
Here $T^2_{\rm dw}(\bfk,z)$ is given by
\ba
T^2_{\rm dw}(\bfk,z) &\equiv& T^2(k) e^{-g_\mu k^2 /(2k_*^2)}
\nonumber \\&&
+T^2_{\rm nw}(k)\left[1-e^{-g_\mu k^2 /(2k_*^2)}\right],
\label{eq:T(k)_dw}
\ea
where $T(k)$ is the linear matter transfer function,
$T_{\rm nw}(k)$ is the pure CDM (no baryons) transfer function given by
\citet[Eq.~29]{EH98}, and
\be
g_\mu(\bfk,z) \equiv G^2(z) \{ 1-\mu^2 +\mu^2 [1+f_g(z)]^2 \}
\label{eq:gmu}
\ee
describes the enhanced damping along the line of sight due to the enhanced power.
The nonlinear damping factor, $e^{-g_\mu k^2 /(2k_*^2)}$, with $g_\mu$ given by Eq.(\ref{eq:gmu}),
was derived by \cite{Eisen07b} using N-body simulations.
Note that since density perturbations grow with cosmic time, the linear regime expands
as we go to higher redshifts.
Hence the function $g_\mu$ scales with the linear growth factor $G(z)$ squared, which corresponds to the
scale of the linear regime increasing with $1/G(z)$ at high redshifts.

The scale $k_*$ is related to the percentage of nonlinearity from \cite{SE07}, $p_{NL}$, via
\be
k_*^{-1}=8.355 \,h^{-1}\mbox{Mpc} \,(\sigma_8/0.8)\, p_{\rm NL}.
\ee
The true galaxy power spectrum should have $p_{\rm NL}=1$. Recently ``reconstruction'' algorithms have been proposed \citep{Eisenstein07} and implemented \citep{Padmanabhan12} that reverse some of the flows and move galaxies back closer to their original (Lagrangian) positions. If such an algorithm is applied to data, the nonlinearity percentage can be reduced. BAO reconstruction is a rapidly developing field, but is in its early stages and high-$z$ redshift surveys may have to deal with survey geometries that are more complex and bias-weighted galaxy densities $b^2n$ that are smaller than that of e.g. BOSS. For the present work, we consider a range of values for $p_{\rm NL}$.
The optimistic case of $p_{\rm NL}=0.5$ corresponds to $k_*\simeq 0.24 \,h/$Mpc,
whereas the most conservative case of $p_{\rm NL}=1$ (no reconstruction) corresponds to $k_* \simeq 0.12 \,h/$Mpc,
assuming $\sigma_8=0.8$.

For an intuitive understanding of the dewiggled power spectrum of Eq.(\ref{eq:P(k|z)}),
we can rewrite its corresponding transfer function, Eq.(\ref{eq:T(k)_dw}), as follows
\ba
T^2_{\rm dw}(\bfk,z)&=& T^2_{\rm nw}(k) + \left[T^2(k)-T^2_{\rm nw}(k)\right] e^{-g_\mu k^2 /(2k_*^2)}\nonumber\\
&\equiv & T^2_{\rm nw}(k) + T^2_{\rm BAO}(k) e^{-g_\mu k^2 /(2k_*^2)}
\ea
where we have defined $T^2_{\rm BAO}(k)=T^2(k)-T^2_{\rm nw}(k)$, the difference between
the linear matter transfer functions with and without baryons. Clearly, the
exponential damping due to nonlinear effects is only applied to the transfer 
function associated with BAO. 
\cite{Angulo08} have compared the spherically-averaged form of this model with measurements 
from simulated data, and found that it works extremely well on the linear and quasilinear scales;
the assessment of its accuracy is presently limited by the shot noise of
currently available numerical simulations.
In future work, we will test this model without spherical averaging using numerical
simulations to fully assess it.
We do not expect this model to continue working well 
on the smallest scales, where the nonlinear damping is coupled with RSD
\citep[e.g.][]{Jennings11,Reid11,CW12b}.

To avoid the direct measurement of the unknown galaxy bias $b(z)$,
we rewrite our model for the measured galaxy power spectrum as \citep{Wang12}
\ba
\label{eq:P(k)fg_scale1}
~~~~~&& \!\!\!\!\!\!\!\!\!\!\!\!\!\!\!\!\!\!\!\!\!\!\!\!
\overline{P_g(k^{\rm ref}_{\perp},k^{ref}_{\parallel})}
\nonumber \\
&\equiv &P_g(k^{\rm ref}_{\perp},k^{ref}_{\parallel})/(h^{-1}\mbox{Mpc})^3 \nonumber\\
&=&\frac{\left[D_A(z)^{\rm ref}\right]^2  H(z)}{\left[D_A(z)\right]^2 H(z)^{\rm ref}}
 \left[\sigma_{g}(z)+ f_g(z)\sigma_{m}(z)\, \mu^2 \right]^2 \nonumber\\
& & \times   \left(\frac{k}{\mbox{Mpc}^{-1}}\right)^{n_s} T^2_{\rm dw}(\bfk,z)
\frac{e^{-k^2\mu^2 \sigma^2_{r,z}}}{1+k^2 \mu^2 \sigma^2_{r,p}/2}  
 + P_{\rm shot},\nonumber\\
\ea
where we have defined
\ba
\sigma_{g}(z) \equiv  b(z)\,G(z) \,\tilde{P}_0^{1/2}
~~~{\rm and}~~~
\sigma_{m}(z) \equiv  G(z)\,\tilde{P}_0^{1/2}.
\ea
The dimensionless power spectrum normalization constant $\tilde{P}_0$ is
just $P_0$ in Eq.~(\ref{eq:P(k|z)}) in appropriate units:
\be
\label{eq:P0til}
\tilde{P}_0 \equiv \frac{P_0}{(\mbox{Mpc}/h)^3 (\mbox{Mpc})^{n_s}}
=\frac{\sigma_8^2}{I_0 \,h^{n_s}}.
\ee
The second part of Eq.~(\ref{eq:P0til}) is relevant if $\sigma_8$ is used to normalize the 
power spectrum. Note that
\be
I_0 \equiv  \int_0^\infty\rd \bar{k}\,  \frac{\bar{k}^{n_s+2}}{2\pi^2}\, 
T^2(\bar{k}\cdot h\mbox{Mpc}^{-1})\,
\left[\frac{3 j_1(8\bar{k})}{8\bar{k}}\right]^2,
\ee
where $\bar{k}\equiv k/[h\,\mbox{Mpc}^{-1}]$, 
and $j_1(kr)$ is spherical Bessel function. 
Note that $I_0=I_0(\omega_m, \omega_b, n_s, h)$. Since $k_{\parallel}$
and $k_\perp$ scale as $H(z)$ and $1/D_A(z)$ respectively, 
$\overline{P_g^{obs}(k)}$ in Eq.(\ref{eq:P(k)fg_scale1}) does not depend on $h$.

Eq.(\ref{eq:P(k)fg_scale1}) is the model we will use in this paper.
Its absorption of the bias factor is analogous to the approach of
\cite{Song09}, who proposed the use of $f_g(z)\sigma_{8}(z)$ to
probe growth of large scale structure.
The difference is that Eq.(\ref{eq:P(k)fg_scale1}) uses $f_g(z)\sigma_{m}(z)
\equiv f_g(z)G(z) \tilde{P}_0^{1/2}$, which does {\it not} introduce an explicit 
dependence on $h$ (as in the case of using $f_g(z)\sigma_{8}(z)$).

\subsection{Parameters and assumptions}

In our method, the full set of parameters that describe the observed $P_g(\bfk)$ are: 
$\{\ln H(z_i)$, $\ln D_A(z_i)$, $\ln[f_g(z_i)\sigma_m(z_i)]$,
$\ln\sigma_{g}(z_i)$, $P_{\rm shot}^i$; $\omega_m$, $\omega_b$, $n_s, k_*, \sigma_z/(1+z)\}$,
where $i$ indicates the $i$-th redshift slice, and 
$\omega_m\equiv \Omega_m h^2$, and $\omega_b\equiv \Omega_b h^2$.
We marginalize over $\{\ln\sigma_{g}(z_i),P_{\rm shot}^i\}$
in each redshift slice, as well as $k_*$ and $\sigma_z/(1+z)$,
to obtain a Fisher matrix for
$\{\ln H(z_i),\ln D_A(z_i),\ln[f_g(z_i)\sigma_m(z_i)]; \omega_m, \omega_b, n_s\}$.
This full Fisher matrix, or a smaller set marginalized over various parameters,
is projected into the standard set of cosmological parameters 
$\{w_0, w_a, \Omega_X, \Omega_k, \omega_m, \omega_b, n_s, \ln A_s\}$.
There are four different ways of utilizing the information from $P(\bfk)$ \citep{Wang12}.

It is important to note that when evaluating the derivatives of $P_g(\bfk^{\rm ref})$
with respect to the parameters described above (required to calculate the Fisher matrix), 
we should \emph{not} extract information from the damping factors due to
systematic uncertainties, in order to adhere to a conservative and robust approach.
These damping factors are only included to represent the loss of information 
at small scales due to nonlinear effects (and, if applicable, redshift uncertainties).
We treat these damping factors as follows when derivatives are taken.

The $g_\mu$ in the nonlinear damping factor, $e^{-g_\mu k^2 /(2k_*^2)}$, is fixed
at fiducial model values when derivatives are taken, to avoid deriving cosmological
information from the NL damping itself. Note that $k$ is scaled as we vary $T_{dw}(\bfk)$
for consistency, and we marginalize over $k_*$ to allow for the significant uncertainty in the NL damping.

The damping factor due to redshift uncertainty, $e^{-k^2\mu^2\sigma_{r,z}^2}$, 
is computed with $\partial r/\partial z$ from the fiducial model, to avoid deriving
cosmological information from the damping itself. We marginalize over $\sigma_z/(1+z)$
to allow for the uncertainty in our knowledge of redshift accuracy.

The RSD factor due to small scale random motion of galaxies,
$1/[1+k^2 \mu^2 \sigma^2_{r,p}/2]$, is fixed at fiducial model values
when derivatives are taken. Note that as we vary $H(z)$, this RSD factor
remains unchanged, since $k\mu=k_\parallel \propto H(z)$, while
$\sigma_{r,p} \propto 1/H(z)$.
The RSD factor is included here to represent the supression of
power due to galaxy peculiar velocities, and not to provide an accurate
modeling of RSD on all scales.

Since our model fits real data well on these scales \citep{CW11},
it represents a step forward in making Fisher matrix forecasting
for galaxy redshift surveys more realistic.

\section{Results}
\label{sec:results}

We will present results on
\be
x_h(z)\equiv H(z) \,s/c ~~~{\rm and}~~~
x_d(z)\equiv D_A(z)/s,
\ee
where $s\equiv r_s(z_d)$ is the sound horizon at the drag epoch, which is
the characteristic scale of BAO.

We assume the fiducial model adopted by the FoMSWG \citep{FoMSWG}, itself
based on the 5-year {\slshape Wilkinson Microwave Anisotropy Probe} (WMAP) results
\citep{WMAP5}:
$\omega_m \equiv \Omega_m h^2=0.1326$, $\omega_b \equiv \Omega_b h^2=0.0227$, 
$h=0.719$, $\Omega_k=0$, $w=-1.0$, $n_s=0.963$, and $\sigma_8=0.798$.

We will first present results for SDSS DR7 LRGs, in order to compare with the results
from actual data analysis. The analysis of current GC data require the assumption
of cosmological priors; we impose the same broad priors as \cite{CW11}.

Next, we will present results for StageIV+BOSS spectroscopic galaxy redshift surveys,
and compare these with those from the previously widely adopted approach derived
from \cite{SE07} and developed in detail in \cite{Wang12}.
No priors are used in deriving $\{x_h(z), x_d(z), f_g(z)\sigma_{m}(z)/s^\alpha\}$
constraints for future surveys, since these provide model-independent 
constraints on the cosmic expansion history
and the growth rate of cosmic large scale structure. These allow
the detection of dark energy evolution, and 
the differentiation between an unknown energy component and modified
gravity as the causes for the observed cosmic acceleration.

In order to derive dark energy figure of merit (FoM), as defined by
the DETF \citep{DETF}, we project our Fisher matrices into the standard 
set of dark energy and cosmological parameters:
$\{w_0, w_a, \Omega_X, \Omega_k, \omega_m, \omega_b, n_s, \ln A_s\}$.
To include Planck priors,\footnote{For a general and robust method for
including Planck priors, see \cite{Mukherjee08}.}
we convert the Planck Fisher matrix for 44 parameters 
(including 36 parameters that parametrize the dark energy equation of state in redshift 
bins) from the FoMSWG into a Planck Fisher matrix for this set of dark energy
and cosmological parameters.

\subsection{Comparison with analysis of data}

To gauge the accuracy of our forecasting methodology compared to the
full analysis of real data, we present our forecasts for 
the SDSS DR7 set of 87,000 LRGs in the redshift range 0.16--0.44 
analyzed by \cite{CW11} in Table \ref{table:sdss}, and compare them with the 
actual measurements performed as part of this work, using
both the SDSS DR7 LRGs and the SDSS LRG mocks from 
LasDamas\footnote{URL: \tt http://lss.phy.vanderbilt.edu/lasdamas/}.

The scale range analyzed by \cite{CW11} is $r=$40-120$\,h^{-1}$Mpc, 
which corresponds to the $k=2\pi/r$ range of 0.0524-0.157$\,h\,$Mpc$^{-1}$.
\cite{CW11} used flat priors on $\omega_b$ and $n_s$ that have widths of $\pm 7\sigma_{WMAP}$
(with $\sigma_{WMAP}$ given by the WMAP seven year results from \cite{Komatsu11}).
In addition, \cite{CW11} imposed flat priors of $0.1< \beta < 0.6$,
$0< \sigma_p <500\,$km/s, and $0.09 < k_*(z=0.35)/[h\,\mbox{Mpc}^{-1}] <0.13$.
We use Gaussian priors on the same parameters with the same widths for the priors as \cite{CW11},
and with means of $\sigma_p=250\,$km/s, and $k_*/[h\,\mbox{Mpc}^{-1}]=0.11 \,G(z=0.33)
=0.0939$ (the width of $k_*/[h\,\mbox{Mpc}^{-1}]$ is $0.02\,G(z=0.33)=0.01707$). 
Note that our definition of $k_*$ is independent of redshift, thus it is divided
by the growth factor at the effective redshift of the data set used by 
\cite{CW11}.\footnote{\cite{CW11} scaled their results from $z_{eff}=0.33$ to
$z_{eff}=0.35$ in order to compare with previous results by other groups.}
In addition, we assume a redshift accuracy of $\sigma\ln (1+z)=5\times 10^{-4}$,
and a bias of $b=2.2$ for the SDSS LRGs. These additional assumptions are not
needed for the analysis of real or mock data, since the overall amplitude is
marginalized over \citep{CW11}.

\begin{table*}
\begin{center}
\begin{tabular}{llcccc}\hline
Model & Method & $x_h(z) $ &  $ x_d(z)$ & $\beta$ & $f_g(z)\sigma_m(z)/s^4$ \\ \hline
\hline
FoMSWG & Fisher matrix  & 7.31\% & 4.99\% & 21.98\% & 20.69\% \\
EuclidRB & Fisher matrix  & 7.09\% & 4.67\% & 22.61\% &  21.05\% \\
\hline
None & MCMC analysis of data  & 5.80\% & 3.74\% & 14.89\% & 14.01\% \\
\hline
None & MCMC analysis of mocks  & 6.64\% & 5.37\%  &  23.72\%  & 22.61 \% \\
\hline
\end{tabular}
\end{center}
\caption{Our Fisher matrix estimate of the percentage precision of measurement of $x_h(z)\equiv H(z) s/c$,
$x_d(z)\equiv D_A(z)/s$, $\beta$,  and $f_g(z)\sigma_m(z)/s^4$ at an
effective redshift of $z=0.35$ from SDSS DR7 LRGs, 
compared to the actual measurements using the anisotropic correlation function per \protect\cite{CW11}.
Eq.(\ref{eq:P(k)fg_scale1}) is used in all cases.} 
\label{table:sdss}
\end{table*}

Since the Fisher matrix results depend on the fiducial model assumed, we
give our Fisher matrix forecasts for two different fiducial models in
Table \ref{table:sdss}: the FoMSWG fiducial model \citep{FoMSWG} with
with $\omega_m \equiv \Omega_m h^2=0.1326$, $\omega_b \equiv \Omega_b h^2=0.0227$, 
$h=0.719$, $\Omega_k=0$, $w=-1.0$, $n_s=0.963$, and $\sigma_8=0.798$, and the Euclid Red Book
fiducial model \citep{RB} with
with $\omega_m \equiv \Omega_m h^2=0.1225$, $\omega_b \equiv \Omega_b h^2=0.021805$, 
$h=0.7$, $\Omega_k=0$, $w=-0.95$, $n_s=0.963$, and $\sigma_8=0.8$.
These fiducial models lead to $\sigma\ln x_h(z)$ and $\sigma\ln x_d(z)$ that differ
by 3.1\% and 6.8\% respectively.

The model used here, Eq.(\ref{eq:P(k)fg_scale1}), differs 
somewhat from that used by \cite{CW11}.
Our new model, Eq.(\ref{eq:P(k)fg_scale1}),
uses {\em aisotropic dewiggling} whereas \citet{CW11} used {\em isotropic dewiggling}, 
which neglects the additional damping along the line of sight due to the enhanced 
Lagrangian displacement in redshift space.\footnote{This is equivalent to setting 
$g_\mu \rightarrow G^2(z)$ in Eq.~(\ref{eq:gmu}).}
Using our Fisher matrix method, we find that assuming isotropic dewiggling
leads to an under-estmate of $\sigma_{\ln x_h(z)}$ and $\sigma_{\ln x_d(z)}$ of 
$\sim 23-24$\% and $\sim 12-13$\% respectively.

In order to make an accurate comparison, we have repeated the analysis of the SDSS DR7 LRG
sample used by \cite{CW11} using Eq.(\ref{eq:P(k)fg_scale1}) as part of this work.
We find that the data and mocks give $\sigma_{\ln x_h(z)}$ and $\sigma_{\ln x_d(z)}$
that differ by 13\% and 30\% respectively, with the mock results agreeing with
our Fisher matrix forecasts at a level of 10\% or better, given the dependence
of the Fisher matrix forecasts on the assumed fiducial model.

Our Fisher matrix forecasts for the measurement uncertainty on $f_g(z)\sigma_m(z)/s^4$ are 
in excellent agreement with the results from the MCMC analysis of the LasDamas SDSS LRG mocks, 
while the results from the MCMC analysis of SDSS DR7 LRG data give significantly smaller
uncertainty on $f_g(z)\sigma_m(z)/s^4$. This is likely due to the apparent excess clustering
of SDSS DR7 LRGs along the line of sight (this is apparent from comparing the mock and
data panels of Fig.1 in \citealt{CW11}), which is likely also responsible for the
smaller than expected measurement uncertainty of $x_h(z)$ and $x_d(z)$ from this data sample.
This excess power along the line of sight explains the widely noted excess power on large scales 
for the spherically-averaged galaxy correlation function (see, e.g., \cite{Cabre09,Kazin10,CWH12}).
Since the BOSS CMASS galaxies do {\it not} have the excess clustering on the same
large scales \citep{Reid12}, the excess large scale clustering of SDSS DR7 LRGs must
be due to sample variance or unknown systematic effects.

Taking all the factors discussed above into consideration, our Fisher matrix forecasts are
in excellent agreement with the results from actual data analysis.
It is reassuring that our Fisher matrix method gives very similar
results compared to actual data analysis, making it a reliable tool
for parameter forecasting for future surveys.

\subsection{Forecasts for a Stage IV galaxy redshift survey}

We perform forecasts for a Stage IV galaxy redshift survey covering an area of 15,000 deg$^2$, using slitless grism spectroscopy to detect the H$\alpha$ emission line. A wavelength range of 1.1--2.0 $\mu$m, corresponding to $0.7<z<2.0$, was assumed. The depth of the survey was computed using instrument parameters (throughput, exposure time, etc.) similar to those provided for the {\slshape Euclid} mission in \cite{RB}; it is thus representative of a next-generation space-based galaxy survey, though it may not correspond precisely to the final {\slshape Euclid} numbers.

All our forecast results are shown for Stage IV plus BOSS.
The BOSS survey is assumed to cover 10,000 (deg)$^2$, a redshift range of $0.1<z<0.7$, with 
a fixed galaxy number density of $n=3\times 10^{-4} h^3$Mpc$^{-3}$, and a fixed linear bias of
$b=1.7$.

We discuss the galaxy yields from a Stage IV galaxy redshift survey in detail in Sec.\ref{sec:ngal}. 
For clustering analysis, we also require the galaxy bias. We use the galaxy bias function for emission line galaxies given by \cite{Orsi10}, which increases with redshift reaching $b=1.7$ at $z=2$. Again, we note that this is likely to be conservative: the recent bias determination of \cite{Geach12} for H$\alpha$ emitters is $b=2.4^{+0.1}_{-0.2}$ at $z=2.23$.
We assume a redshift accuracy of $\sigma_z/(1+z)=0.001$, and a peculiar velocity dispersion of $\sigma_p=290\,$km/s.

We consider two different cutoffs in scale: $k_{max}=0.2\,h$/Mpc and $k_{max}=0.2\,h$/Mpc,
in order to include the quasilinear regime only in our forecasts.
The choice of $k_{max}=0.2\,h$/Mpc is conservative, and represents the lower bound of the scale range
in which our model works well in analyzing real data.
The choice of $k_{max}=0.3\,h$/Mpc is more optimistic, but represents a feasible goal for 
the lower bound of the scale range in which future studies will enable robust and accurate modeling.

\subsubsection{Galaxy yields for a Stage IV redshift survey}
\label{sec:ngal}

Galaxy yields were computed using the exposure time calculator described in \cite{Hirata12}. Two exposures on each field in each grism bandpass were assumed. The zodiacal background was set to that at 45$^\circ$ ecliptic latitude and 90$^\circ$ away from the Sun at the mean of the annual cycle, and we include a foreground dust column of $E(B-V)=0.05$ magnitudes; these values vary over any realistic survey but are representative. Standard read noise assumptions for the 2k$\times$2k Teledyne HgCdTe detectors were used (32 channel readout, 1 frame per 1.3 s, 20 electrons rms per correlated double sample, with a noise floor of 5 electron rms for many reads). The galaxy survey was assumed to be 70 per cent complete down to the flux limit for a $7\sigma$ significance matched-filter detection.\footnote{Note that some forecasts in the literature use other definitions of detection significance, based on other extraction apertures. The differences are often tens of percents and occasionally as large as a factor of 2. 
The matched-filter method gives the highest reported significance.} The extinction-corrected H$\alpha$ flux limit varies with redshift and galaxy size, but is in the range of $(2.2-3.6)\times 10^{-16}$ erg$\,$s$^{-1}$cm$^{-2}$ for a source half-light radius of 0.3 arcsec.

As a point of comparison, we ran the \cite{Hirata12} code on the Hubble Space Telescope Wide Field Camera 3 (HST/WFC3) G141 grism ($1.1<\lambda<1.7$ $\mu$m), using parameters from the Instrument Handbook \citep{Dressel11}. We find that for an exposure time of 2700 s, the $5\sigma$ sensitivity should be $(4.6-8.2)\times 10^{-17}$ erg$\,$s$^{-1}$cm$^{-2}$, with the lower (better) numbers at the red end of the bandpass. This is in good agreement ($\sim 20$ per cent) with the median sensitivity actually achieved by WFC3 observations -- see e.g. Figure 5 of \cite{Atek10}.

The line flux sensitivity and completeness are only part of determining the number of redshifts obtained by a survey -- one also needs a luminosity function. In the past decade of space-based redshift survey mission planning, the H$\alpha$ luminosity function (H$\alpha$LF) has been a matter of vigorous debate: direct measurements have suffered from small-number statistics, while indirect methods (based on scaling from rest-frame ultraviolet or [O~{\sc ii}] luminosities) have had difficult-to-quantify systematic errors. Indeed, the estimates used for space mission planning \citep[e.g.][]{Yan99,Hopkins00,Reddy08,Jouvel09,Geach10,Sobral12} have spanned a factor of $\sim 3$ in number density, even accounting for the different cosmologies assumed. Fortunately, empirical measures of the H$\alpha$LF across the relevant range of redshifts with large-number statistics (dozens of objects in the relevant flux range, in multiple fields) are now available.

We use the H$\alpha$LF of \cite{Sobral12}, with conversions described in \cite{Hirata12} \S3E to ensure consistency with the exposure time calculator inputs.
This is based on blind narrow-band surveys, and updates
the previous estimate by \cite{Geach10}. The
new H$\alpha$LF is lower than the previous estimate; in approximate decreasing order of importance, the main differences are:
\newcounter{lf}
\begin{list}{\arabic{lf}.}{\usecounter{lf}}
\item Consistent treatment of internal (host galaxy) extinction corrections, which are applied to some
H$\alpha$LF results and must be undone to predict redshift
survey yields.
\item Improved statistics and addition of data at new redshifts.
\item Redshift-averaging effects in some of the grism luminosity functions (this does not occur in narrowband
surveys).
\item Aperture corrections. 
\item Conversion to the WMAP-5/FoMSWG cosmology.
\end{list}
The narrowband surveys do not cleanly separate the H$\alpha$ 6563 \AA\ line from the [N~{\sc ii}] doublet at 6548,6583 \AA, and the \cite{Sobral12} H$\alpha$LF removes the estimated [N~{\sc ii}] contribution. Of course, in a grism survey the two lines will be a partial blend, thus we may be underestimating the final detection significance of the galaxies. For this reason, we expect that our analysis is somewhat conservative.

Table \ref{table:ngal} gives our resultant galaxy yields as a function of redshift.

\begin{table*}
\begin{center}
\begin{tabular}{ccccccc}
\hline
            $z$  & $\lambda$  & EE50  & d$V$/(d$z\cdot$ d$A$)   & F$_{lim}$@0.30$^{\prime\prime}$ &  $n$   & d$N$/d$z\cdot$d$A$  \\
                 & ($\mu$m) & (arcsec) & (Mpc$^3$ deg$^{-2}$) &       (W m$^{-2}$)&   (Mpc$^{-3}$)  & (deg$^{-2}$) \\
\hline
	   0.700 &1.1160 &0.2297 &5.56334E+06 &3.56250E$-$19 &1.70876E$-$04 &9.50640E+02\\
 	   0.750 &1.1489 &0.2315 &6.06057E+06 &3.33879E$-$19 &1.89962E$-$04 &1.15128E+03\\
 	   0.800 &1.1817 &0.2335 &6.54313E+06 &3.15425E$-$19 &2.11061E$-$04 &1.38100E+03\\
 	   0.850 &1.2145 &0.2354 &7.00894E+06 &3.01278E$-$19 &2.23002E$-$04 &1.56301E+03\\
 	   0.900 &1.2474 &0.2373 &7.45644E+06 &2.89615E$-$19 &2.01417E$-$04 &1.50185E+03\\
 	   0.950 &1.2802 &0.2393 &7.88451E+06 &2.79942E$-$19 &1.81900E$-$04 &1.43419E+03\\
 	   1.000 &1.3130 &0.2413 &8.29240E+06 &2.72837E$-$19 &1.62666E$-$04 &1.34889E+03\\
 	   1.050 &1.3458 &0.2433 &8.67969E+06 &2.67418E$-$19 &1.44931E$-$04 &1.25796E+03\\
 	   1.100 &1.3787 &0.2454 &9.04621E+06 &2.63267E$-$19 &1.28950E$-$04 &1.16651E+03\\
 	   1.150 &1.4115 &0.2474 &9.39200E+06 &2.60703E$-$19 &1.13930E$-$04 &1.07003E+03\\
 	   1.200 &1.4443 &0.2494 &9.71731E+06 &2.64718E$-$19 &9.39210E$-$05 &9.12659E+02\\
 	   1.250 &1.4771 &0.2696 &1.00225E+07 &2.80452E$-$19 &6.69603E$-$05 &6.71109E+02\\
 	   1.300 &1.5100 &0.2714 &1.03081E+07 &2.68660E$-$19 &6.78461E$-$05 &6.99362E+02\\
 	   1.350 &1.5428 &0.2731 &1.05746E+07 &2.59173E$-$19 &6.78039E$-$05 &7.16996E+02\\
 	   1.400 &1.5756 &0.2749 &1.08226E+07 &2.51215E$-$19 &6.73077E$-$05 &7.28446E+02\\
 	   1.450 &1.6084 &0.2768 &1.10529E+07 &2.44917E$-$19 &6.61038E$-$05 &7.30641E+02\\
 	   1.500 &1.6413 &0.2786 &1.12662E+07 &2.39935E$-$19 &6.34722E$-$05 &7.15093E+02\\
 	   1.550 &1.6741 &0.2804 &1.14632E+07 &2.35608E$-$19 &6.03252E$-$05 &6.91521E+02\\
 	   1.600 &1.7069 &0.2823 &1.16446E+07 &2.31880E$-$19 &5.73065E$-$05 &6.67313E+02\\
 	   1.650 &1.7397 &0.2841 &1.18112E+07 &2.29092E$-$19 &5.40911E$-$05 &6.38881E+02\\
 	   1.700 &1.7726 &0.2860 &1.19637E+07 &2.26675E$-$19 &5.11258E$-$05 &6.11651E+02\\
 	   1.750 &1.8054 &0.2879 &1.21027E+07 &2.24586E$-$19 &4.83888E$-$05 &5.85637E+02\\
 	   1.800 &1.8382 &0.2898 &1.22291E+07 &2.22811E$-$19 &4.58519E$-$05 &5.60729E+02\\
 	   1.850 &1.8710 &0.2917 &1.23435E+07 &2.21390E$-$19 &4.34464E$-$05 &5.36281E+02\\
 	   1.900 &1.9039 &0.2936 &1.24465E+07 &2.20333E$-$19 &4.11589E$-$05 &5.12286E+02\\
 	   1.950 &1.9367 &0.2955 &1.25388E+07 &2.20230E$-$19 &3.86029E$-$05 &4.84035E+02\\
 	   2.000 &1.9695 &0.2975 &1.26210E+07 &2.20856E$-$19 &3.59559E$-$05 &4.53799E+02\\
\hline
\end{tabular}
\end{center}
\caption{Galaxy yields for a 2-exposure Stage IV galaxy redshift survey as discussed in Sec.~\ref{sec:ngal}.
Columns indicate: the redshift; observer-frame wavelength of H$\alpha$; the half-light radius (encircled energy 50\%, EE50) of the point-spread function; the cosmological volume element $\rd V/(\rd z\cdot\rd A)$; the limiting flux for a 0.3 arcsec half-light radius galaxy; the number $n$ of observed sources per unit comoving volume; and the number of sources d$N$/d$z\cdot$d$A$ per unit redshift per unit solid angle.}
\label{table:ngal}
\end{table*}

\subsubsection{Dark energy figure of merit results}

Table \ref{table:stageIV_new} shows the dark energy figure-of-merit (FoM) \citep{Wang08b},
\be
FoM(w_0, w_a) \equiv \frac{1}{\sqrt{\det\mbox{Cov}(w_0,w_a)}}
\ee
for the four different approaches to utilizing the information from the measured
anisotropic galaxy power spectrum \citep{Wang12}: \\
\noindent
(1) $\{x_h(z), x_d(z)\}$ from $P(\bfk)$;\\
\noindent
(2) $\{x_h(z), x_d(z), f_g(z) \sigma_m(z)/s^\alpha\}$ from $P(\bfk)$;\\
\noindent
(3) $P(\bfk)$, marginalized over $f_g(z) \sigma_m(z)$;\\
\noindent
(4) $P(\bfk)$+$f_g(z)$; $P(\bfk)$ including $f_g(z) \sigma_m(z)$.

\begin{table*}
\begin{center}
\begin{tabular}{cccccccc}\hline
$k_{max}$ & $k_*$ & FoM & FoM$_{GR}$ $\,\,\,\,$(FoM, d$\gamma$) &  FoM &  FoM$_{GR}$ $\,\,\,\,$(FoM, d$\gamma$)\\
($h\,$Mpc$^{-1}$) & ($h\,$Mpc$^{-1}$) & $\{x_h,x_d\}$ &$\{x_h,x_d,f_g\sigma_m/s^\alpha\}$ &  $\{P(\bfk)\}$ & $\{P(\bfk)$+$f_g\}$\\
\hline
0.2  &   0.12  &    6.56  &   30.78   \hskip 0.1in   (22.80, $\,$0.0514)  & 14.81  &   40.48  \hskip 0.1in   (24.47, $\,$0.0476)\\
0.2  &   0.24  &    10.05  &   45.30  \hskip 0.1in   (31.07, $\,$0.0470)  & 23.37  &   54.59  \hskip 0.1in   (35.05, $\,$0.0456)\\
0.3 &    0.12  &    9.83 &    44.11  \hskip 0.1in    (33.98, $\,$0.0437)  & 19.94  &   65.32   \hskip 0.1in  (35.19, $\,$0.0386)\\
0.3 &    0.24 &    12.73   &  62.51  \hskip 0.1in    (42.41, $\,$0.0394)  & 29.81  &   79.57  \hskip 0.1in   (46.43, $\,$0.0374)\\
\hline
 &&&Stage IV+BOSS+Planck & &&&\\
\hline
0.2  & 0.12 &    58.30 &   139.24   \hskip 0.1in   (80.49, $\,$0.0392) &  61.30   &   171.90 \hskip 0.1in   (106.98, $\,$0.0341)\\
0.2 &  0.24 &    92.64  &  193.61   \hskip 0.1in   (119.58, $\,$0.0376)  & 96.22   &   238.63 \hskip 0.1in  (152.10, $\,$0.0329)\\
0.3  & 0.12 &    85.24 &   209.98   \hskip 0.1in   (110.10, $\,$0.0344) & 89.29   &   240.11  \hskip 0.1in  (136.75, $\,$0.0311)\\
0.3 &  0.24 &  119.88	&273.20	\hskip 0.1in       (152.31, $\,$0.0322)&  123.55   &   315.23 \hskip 0.1in  (184.21, $\,$0.0295)\\
\hline
\end{tabular}
\end{center}
\caption{Our Fisher matrix forecasts for Stage IV+BOSS galaxy redshift surveys using our new galaxy power spectrum model, 
Eq.(\ref{eq:P(k)fg_scale1}), for the four cases discussed
in \protect\cite{Wang12}. FoM$_{GR}$ denoted the FoM assuming general relativity. The parameter
$\gamma$ is defined by $f_g(z)=[\Omega_m(a)]^\gamma$.} 
\label{table:stageIV_new}
\end{table*}

It is clear from Table \ref{table:stageIV_new} that for a given cutoff $k_{max}$, the FoM for $(w_0,w_a)$
increases as we increase the dewiggling scale $k_*$ (i.e., decrease the nonlinear effects).
For a fixed level of nonlinearity (i.e., fixed $k_*$), the FoM for $(w_0,w_a)$ increases as
we increase the cutoff $k_{max}$.
The scaling of $f_g(z) \sigma_m(z)$ with $s$ depends on the level of nonlinearity
assumed: for 50\% nonlinearity ($k_*=0.24\,h/$Mpc), $\alpha \simeq 4$, while
for 100\% nonlinearity ($k_*=0.12\,h/$Mpc), $\alpha \simeq 5$. This is not surprising,
since the scaling of $f_g(z) \sigma_m(z)$ with $s^4$ (i.e., $\alpha=4$) originates
from the linear matter power spectrum \citep{Wang12}. When nonlinear effects are
fully included (and not assumed to be reduced due to density field reconstruction),
the appropriate model for $P(\bfk)$ (i.e., Eq.[\ref{eq:P(k)fg_scale1}]) deviates significantly
from the linear power spectrum, leading to modification of the scaling of
$f_g(z) \sigma_m(z)$ with $s$.

The choice of $\alpha$ does not affect the FoM($w_0,w_a)$ from the $\{x_h, x_d, f_g\sigma_m/s^\alpha\}$
case, as the correlations between $f_g(z)\sigma_m(z)/s^\alpha$ and $\{x_h(z), x_d(z)\}$
depend on $\alpha$. Choosing the $\alpha$ that minimizes the uncertainties in
$f_g(z)\sigma_m(z)/s^\alpha$ does maximize the FoM($w_0,w_a)$ when Planck priors
are included.

\subsection{Comparison with previous work}

Previously, the forecasts of dark energy constraints from full $P(\bfk)$ assumed that
\ba
P_{g}^{old}(k^{ref}_{\perp},k^{ref}_{\parallel}) &=&
\frac{\left[D_A(z)^{ref}\right]^2  H(z)}{\left[D_A(z)\right]^2 H(z)^{ref}}
\nonumber \\ && \times
b^2 \left( 1+\beta\, \mu^2 \right)^2 P_{lin}(\bfk|z)
 \nonumber\\
&& \times e^{-\frac{1}{2}k^2 \Sigma_{nl}^2 }e^{-k^2\mu^2 \sigma^2_{r;z,p}} + P_{\rm shot},
\label{eq:P(k)old}
\ea
where the linear matter power spectrum $P_{lin}(\bfk|z)= G^2(z) P_0 k^{n_s} T^2(k)$ (with
$T(k)$ denoting the linear matter transfer function), and
\be
\sigma^2_{r;z,p} = \left(\frac{\partial r}{\partial z}\right)^2\,\left[\sigma_z^2 +
\left(\frac{\sigma_p}{c}\right)^2\right]
\ee
Alternatively, we can write
\ba
~~~~~~
&& \!\!\!\!\!\!\!\!\!\!\!\!\!\!\!\!\!\!\!\!\!\!\!\!
\overline{P_g^{old}(k^{ref}_{\perp},k^{ref}_{\parallel})} \nonumber\\
&\equiv &P_g^{old}(k^{ref}_{\perp},k^{ref}_{\parallel})/(h^{-1}\mbox{Mpc})^3 \nonumber\\
&=&\frac{\left[D_A(z)^{ref}\right]^2  H(z)}{\left[D_A(z)\right]^2 H(z)^{ref}}
 \left[\sigma_{g}(z)+ f_g(z)\sigma_{m}(z)\, \mu^2 \right]^2 \nonumber\\
& & \times   \left(\frac{k}{\mbox{Mpc}^{-1}}\right)^{n_s} T^2(k)\,
e^{-\frac{1}{2}k^2 \Sigma_{nl}^2 }e^{-k^2\mu^2 \sigma^2_{r;z,p}}
\nonumber \\ &&
 + P_{\rm shot}.
\label{eq:P(k)fg_scale1_old}
\ea

Table \ref{table:stageIV_old} lists the FoM for $(w_0,w_a)$ for the same four cases as
listed in Table \ref{table:stageIV_new}. Each line in Table \ref{table:stageIV_old}
and its corresponding line in Table \ref{table:stageIV_new} assume the {\emph same}
level of nonlinearity and the same cutoff $k_{max}$. The only difference between
the two tables is the model assumed for $P(\bfk)$: Eq.~(\ref{eq:P(k)fg_scale1})
(from Eq.~[\ref{eq:P(k)b}]) is assumed for Table \ref{table:stageIV_new}, 
while Eq.~(\ref{eq:P(k)fg_scale1_old}) (from Eq.~[\ref{eq:P(k)old}]) is assumed for Table \ref{table:stageIV_old}. 

Note that the two assumed models of $P(\bfk)$ give similar FoM for all the cases that 
marginalize over the growth information, and for the cases that include growth information
but assume only a nonlinearity level of 50\% ($p_{NL}=0.5$ or $k_*=0.24 \,h/$Mpc).
When we assume a nonlinearity level of 100\% ($p_{NL}=1$ or $k_*=0.12 \,h/$Mpc),
our new model (Eq.~[\ref{eq:P(k)fg_scale1}]) gives significantly larger FoM for the
cases that include the growth information. This is because the old model in
Eq.~(\ref{eq:P(k)fg_scale1_old}) simply damps the linear matter power spectrum exponentially, 
while the new model in Eq.~(\ref{eq:P(k)fg_scale1}) only damps the BAO oscillations, and retain
smaller scale information via the ``no-wiggle'' matter power spectrum
$P_{nw}(\bfk|z)= G^2(z) P_0 k^{n_s} T^2_{nw}(k)$ (with $T_{nw}(k)$ denoting
the zero baryon transfer function from \cite{EH98}).
This results in significantly smaller uncertainties in $\ln\beta(z)$ (and
$\ln f_g(z) \sigma_m(z)/s^\alpha$) when our new model is used, which
in turn leads to significantly larger FoM($w_0,w_a)$ when growth information
is included.

While the scaling of $f_g(z) \sigma_m(z)$ with $s$ depends on the level of nonlinearity
assumed in our new model (see discussion in the previous subsection),
$f_g(z) \sigma_m(z)$ scales with $s^4$ in the old model \citep{Wang12}.
We find that our FoM results are not sensitive to the exact choice of $\alpha$;
we have chosen $\alpha=4$ for all the FoM tabulated in Tables \ref{table:stageIV_new}
and \ref{table:stageIV_old} when growth information is included.

\begin{table*}
\begin{center}
\begin{tabular}{cccccccc}\hline
\hline
$k_{max}$ &$p_{NL}$ & FoM &FoM$_{GR}$ $\,\,\,\,$(FoM, d$\gamma$) &  FoM & FoM$_{GR}$ $\,\,\,\,$(FoM, d$\gamma$)\\
($h\,$Mpc$^{-1}$) & & $\{x_h,x_d\}$ &$\{x_h,x_d,f_g\sigma_m/s^\alpha\}$ &  $\{P(\bfk)\}$ & $\{P(\bfk)$+$f_g\}$\\
\hline
0.2 &  1.0   &    5.59  &   21.59  \hskip 0.1in    (14.93, $\,$0.0703)  & 13.95   &  27.35  \hskip 0.1in    (18.05, $\,$0.0625)\\
0.2&   0.5  &    11.70  &   46.05  \hskip 0.1in    (32.66, $\,$0.0498)  & 26.58  &   55.96   \hskip 0.1in    (36.89, $\,$0.0463)\\
0.3 &  1.0    &   5.91  &   23.45   \hskip 0.1in    (15.79, $\,$0.0686) &  15.51 &    30.74    \hskip 0.1in    (19.67, $\,$0.0603)\\
0.3 &  0.5  &    14.28  &   59.20  \hskip 0.1in    (40.79, $\,$0.0452)  & 35.05  &   76.51    \hskip 0.1in    (47.19, $\,$0.0416)\\
\hline
&&& Stage IV+BOSS+Planck &&&&\\
\hline
0.2  & 1.0   &   53.03 &   105.76      \hskip 0.1in    (72.57, $\,$0.0503) &  53.61  &  132.76   \hskip 0.1in    (101.98, $\,$0.0442)\\
0.2 &  0.5  &   109.03 &   211.49     \hskip 0.1in    (134.32, $\,$0.0393) & 110.39  &  247.21   \hskip 0.1in    (167.93, $\,$0.0360)\\
0.3  & 1.0  &    56.80  &  114.70      \hskip 0.1in    (78.48, $\,$0.0489) &  57.57  &  142.39   \hskip 0.1in    (108.80, $\,$0.0431)\\
0.3&   0.5  &   134.30&    270.66     \hskip 0.1in    (166.07, $\,$0.0361) & 136.51  &  307.96   \hskip 0.1in    (199.79, $\,$0.0336)\\
\hline
\end{tabular}
\end{center}
\caption{Our Fisher matrix forecasts for Stage IV+BOSS galaxy redshift surveys using the galaxy power spectrum model from previous
work, Eq.(\ref{eq:P(k)fg_scale1_old}), for the four cases discussed in \protect\cite{Wang12}.
FoM$_{GR}$ denoted the FoM assuming general relativity. The parameter
$\gamma$ is defined by $f_g(z)=[\Omega_m(a)]^\gamma$.} 
\label{table:stageIV_old}
\end{table*}

\section{Conclusion}
\label{sec:conclusion}

We have shown that the forecasting of dark energy constraints from galaxy redshift surveys
can be improved in fidelity by using the ``dewiggled'' galaxy power spectrum, $P_{\rm dw}(\bfk)$, 
in the Fisher matrix calculations. Since $P_{\rm dw}(\bfk)$ is a good fit to real galaxy clustering data over most
of the scale range of interest, our approach is more realistic compared to
previous work in forecasting dark energy constraints from galaxy redshift surveys.

We tested our methodology by comparing our Fisher matrix forecasts 
with results from actual data analysis, and found excellent agreement (see Table \ref{table:sdss}).
Our Fisher matrix method gives very similar results compared to actual data analysis, 
making it a reliable tool for parameter forecasting for future surveys.

Using our new approach, we studied a Stage IV galaxy redshift survey, in combination with
BOSS, without and with Planck priors.
We find that in this new approach, increasing nonlinear effects from 50\%
(best case) to 100\% (most conservative) has a significantly reduced 
impact on the dark energy figure of merit compared to previous work. This indicates
that the erasure of information by nonlinear smearing is having only a modest effect 
on our ability to constrain cosmology using the ``full P(k)'' method
in our new realistic approach.

\section*{Acknowledgments}

Y.W. was supported in part by DOE grant DE-FG02-04ER41305.
C.C. was supported by the Spanish MICINN’s Consolider-Ingenio 2010 Programme under grant MultiDark CSD2009-00064 and grant AYA2010-21231.
C.H. was supported by DOE DOE.DE-SC0006624, and the David and Lucile Packard Foundation.

\setlength{\bibhang}{2.0em}

\label{lastpage}

\end{document}